# Deep Nonlinear Hyperspectral Unmixing Using Multi-task Learning

Saeid Mehrdad[1], Seyed AmirHossein Janani[2]

[1] Department of Computer Science and Engineering, University of Notre Dame, Notre Dame, IN 46556 USA

[2] School of Electrical Engineering, Iran University of Science &Technology, Tehran 16846-13114, Iran

*Abstract*— In hyperspectral images, due to the low spatial resolution of hyperspectral sensors, constituent materials of a scene can be mixed because of their spatial interactions. Therefore, hyperspectral unmixing – decomposing mixed pixels into a set of endmembers and abundance fractions – is an important task in hyperspectral image processing. Nonlinear spectral unmixing has recently received considerable attention, since there are many situation in which a linear mixture model is not appropriate and a nonlinear model is preferable. In this paper, we propose an unsupervised nonlinear unmixing approach that, in contrast to most of the existing nonlinear unmixing methods which are based on specific assumptions on the nonlinearity, introduces a general nonlinear model by using the potential of deep learning in solving nonlinear problems. This model is a two-branch deep neural network. In the first branch, the endmembers are learned by reconstructing the rows of the hyperspectral image through a series of hidden neural network layers. In the second branch, the abundance fractions are learned by doing a similar process for columns of the image. Based on multi-task learning, we introduce an auxiliary task which enforces the two branches to work together. Multi-task learning improves the performance of the two branches and can be considered as a regularizer mitigating overfitting. Extensive experiments with synthetic and real data verify the effectiveness of the proposed method compared with state-of-the-art hyperspectral unmixing methods.

## I. Introduction

HYPERSPECTRAL remote sensing images, gathered by spectral imaging sensors in hundreds or thousands of contiguous spectral bands, provide spatial and spectral information used in a wide range of applications including, environmental monitoring, agriculture, land cover classification, and spectromicroscopy, to name but a few [1-5]. However, because of such factors as the relatively low spatial resolution of imaging devices and their long distance to the targets of monitoring, the diversity of objects in a scene under study, and multiple scattering [1], the pixels of a hyperspectral image may be mixed by several pure materials, leading to difficulties to the analysis and characterization of hyperspectral data [6]. Therefore, hyperspectral unmixing is an important task to deal with mixed pixels. In hyperspectral unmixing methods, the spectrum of mixed pixels is decomposed into a set of pure constituent spectra, termed endmembers, representing the pure materials, and a collection of fractional abundances, representing the percentage of each endmember [1]. According to the approaches, it is possible to obtain these two unknown variables simultaneously or individually.

In Hyperspectral unmixing methods, either a linear mixing model (LMM) or a nonlinear mixing model is considered for the mixture of spectra in mixed pixels [7]. In linear models, the spectral reflection in an observed pixel is assumed as a linear combination of the spectral signatures of the endmembers with appropriate weights according to their corresponding abundance proportion. Due to its simplicity and physical interpretation, the linear mixing model has received a vast amount of attention and research in the signal and image processing and geoscience literatures [8-13]. These models provide promising results for some real-world scenes such as flat landscape and irradiance homogeneity in the observed scene [7]. However, in such conditions as multiple scattering effects [14], microscopic-level material mixtures [15], and water-absorbed environments [16], the incident light interacts in a complex manner among the endmembers in the scene, which leads to higher-order photon interactions that cause nonlinear effects in the spectrum of the mixed pixels. As a consequence, the linear mixing model loses its accuracy under this situation. Thus, a nonlinear mixing model is required to provide a more accurate mixing model for the mixed pixels [17-19], as we adopted in this paper.

A considerable amount of research has been conducted to address the nonlinear mixing model. For instance, bilinear models and their extension are widely used to take into account the nonlinear effects [20-26]. To provide a better approximation for a wide area of second-order nonlinearities, a polynomial post-nonlinear model is applied in [27]. Adding a supplementary term to the linear mixing model that addresses particular nonlinear effects is a common feature in the above models. This additive term is mainly described by a function defined in a reproducing kernel Hilbert space [17]. Spatial regularization [28] and neighborhood dependent contributions [29] are also applied as further extensions of this model. Using a ray-based approximation of light and a graph-based description of the optical interactions, a nonlinear mixing model is presented in [30]. Nevertheless, a requirement for considering a specific form of nonlinearity, which restricts the generalization of the model, is a major drawback for the models mentioned above. In addition, choosing appropriate kernels and its parameters is a non-trivial problem that imposes a limitation on the applications of kernel-based approaches providing models with flexible nonlinear terms.

In recent years, because of its promising performance in computer vision, natural language processing, and image classification, deep neural networks have been employed in several remote sensing applications [31-34]. However, as compared to other applications in remote sensing, deep learning has not been applied as extensively to the hyperspectral unmixing problem. Among the deep learning-based approaches for hyperspectral unmixing, a convolutional neural network has been applied in [35]; however, requiring a training set with known abundance ground-truths or endmembers is a drawback for this approach. In [36], an end-to-end deep neural network model that has an alternating architecture and uses both model-based methods and learning-based methods is introduced for hyperspectral unmixing. Another deep learning-based algorithm for hyperspectral unmixing in which the endmembers are extracted by employing a simplex volume maximization, and abundances are estimated by applying a deep image prior has been presented in [37]. Additionally, autoencoder, a type of neural network architecture compressing an input into a lower-dimensional space which can be uncompressed to reconstruct the original input, has demonstrated a great performance in estimating both endmembers and abundances in the hyperspectral unmixing problem [38-43]. Apart from these, convolutional neural network-based approaches have also received attention in extracting spectral features from hyperspectral images [44, 45]. Despite the great potential of deep learning in solving a variety of nonlinear problems, the aforementioned approaches are designed to address the linear mixture model and fail to employ such potential. In addition, in deep neural networks, particularly autoencoder-based algorithms – especially when the training data is sparse – overfitting is a potential problem that should be considered when these approaches are used.

In this paper, we proposed a new deep learning-based framework for nonlinear hyperspectral unmixing. The intuition behind our method is that the endmembers and abundance fractions are the latent representations of the rows and columns of the hyperspectral image, respectively. Therefore, in order to obtain these latent representations, we design a two-branch deep neural network. The first branch receives unknown latent variables of rows, endmembers, and tries to reconstruct the rows of the hyperspectral image with them. It reconstructs the nonlinear interactions among the spectrum of the image through a series of hidden neural network layers via nonlinear activation functions. In the meantime, the weights of the network and the input are enforced to reconstruct the linear effects in parallel with the nonlinear structure. By optimizing the input and the parameters of the network, the reconstruction error between the rows of the real image and the output of the network is minimized. By doing so, the structure is trained to obtain the proper nonlinear latent representations of the rows, i.e. endmembers. In order to obtain the latent representations of the columns, abundance fractions, the same process is done in the second branch to reconstruct the columns of the hyperspectral image. By utilizing multi-task learning principles, we employ the obtained latent representations of rows and columns to construct a nonnegative matrix factorization (NMF) model as a related task, which enforces the two separated branches to work together. This task performs as a regularizer, which not only improves the performance of the unmixing task but also mitigates overfitting. To evaluate the performance of the proposed method and compare it with the state-of-the-art methods of hyperspectral unmixing, several experiments have been conducted on synthetic and real data. The experimental results show that the proposed method outperforms other methods in extracting endmembers and abundance fractions. We also evaluate the influence of the proposed regularization technique on the accuracy of the estimated endmembers and abundance fractions. It is important to note that the proposed neural network differs from conventional neural networks in some ways. While the inputs of conventional neural networks are known, those of our method are unknown and must be optimized.

In the rest of the paper, we have arranged the content as follows. In Section II, the formulation of the proposed approach is presented. In Section III, the proposed approach is described in detail. In section IV experimental results are presented, and finally, in Section V, we conclude our paper.

## II. PROBLEM FORMULATION

Given an observation matrix $X \in \mathbb{R}^{P \times N}$, in which $P$ is the number of spectral bands and $N$ denotes the number of pixels, spectral unmixing is to decompose $X$ into a set of individual $K$ endmember $E \in \mathbb{R}^{P \times K}$ which has the endmembers as columns and estimating their corresponding abundance fractions matrix $A \in \mathbb{R}^{K \times N}$. The formulation of linear mixing model is described in

$$x_n = \sum_{k=1}^{K} a_{k,n} e_k + n_n = E a_n + n_n, \quad (1)$$

where $x_n$ denotes the spectra of pixel number $n$ and $n = 1, \dots, N$, $e_k = [e_{1k}, \dots, e_{Pk}]^T$ denotes the $k$-th endmember spectrum, $a_n = [a_{1n}, \dots, a_{Kn}]^T$ is the abundance vector for pixel $n$, and $n_n$ is the observation noise.

The abundance fraction values should naturally meet two physical constraints, abundance nonnegativity constraint (ANC) and abundance sum-to-one constraint (ASC). These two constraints are formulated as follows:

$$\forall\, k, n:\ a_{k,n} \geq 0 \quad (2)$$

$$\sum_{k=1}^{K} a_{k,n} = 1 \quad (3)$$

By considering the whole image at once, the matrix formulation of Eq. (1) can be written as:

$$X = EA + N \quad (4)$$

In order to consider the nonlinear interactions among the endmembers, we introduce an unmixing model which is a combination of a linear and nonlinear latent variable models. In our work, as endmembers are the latent variables of rows, the rows of the hyperspectral data are considered as a combination of linear and nonlinear transposition of endmembers:

$$x_i^r = \ell(E) + \varphi(E), \quad (5)$$

where $x_i^r$ is the $i$-th row of the data, $\ell(.)$ is a linear function, and $\varphi(.)$ is a nonlinear function.

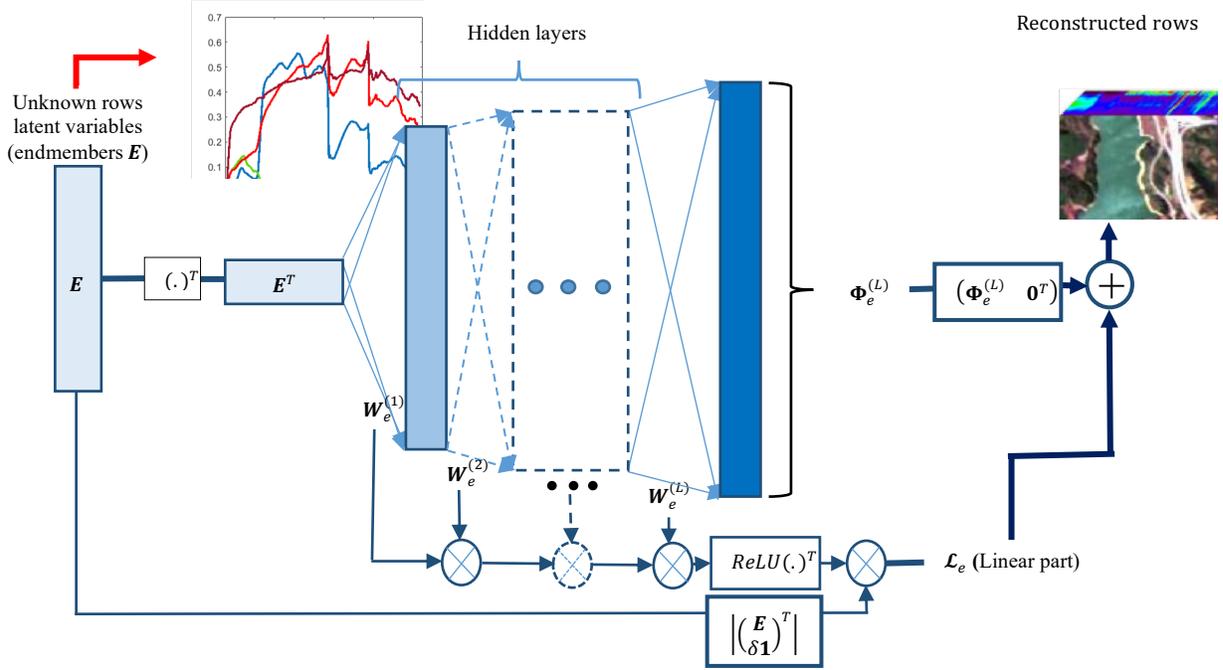

Fig. 1: The diagram of the first branch of the proposed method. The network is trained by reconstructing the rows of the data. After training, the enmembers matrix $E$ is obtained. By satisfying ANC and ASC, $ReLU\left(\prod_{l=1}^{L}(W_e^{(l)})\right)^T$ can be considered as the abundance fraction matrix.

Similarly, as abundances are the latent variables of columns, the columns of the hyperspectral data are considered as a combination of linear and nonlinear transposition of the abundance fractions:

$$x_i^c = \ell(A) + \varphi(A), \quad (6)$$

where $x_i^c$ is the $i$-th column of the data.

In this paper, the problem of hyperspectral unmixing is solved using a two-branch deep neural network. The first branch is designed to estimate the endmember matrix $E$, using the introduced models in Eq. (5), and the second one is designed to estimate the abundance fraction matrix $A$, using the introduced models in Eq. (6). Then, based on multi-task learning principles, we employ the learned latent variables of the two branches, $E$ and $A$, to realize the model in Eq. (4). By doing so, the two branches are forced to work together. The problem is interpreted as a blind unmixing problem and is solved in an unsupervised manner.

## III. PROPOSED APPROACH

In this section the proposed nonlinear unmixing method is described in detail.

### A. The estimation of endmembers

The first branch takes the unknown latent variables $E$ as inputs and reconstructs the rows of the hyperspectral data with a combination of linear and nonlinear transposition of endmembers as the output. In this branch, the nonlinear part of the Eq. (5) is approximated through some hidden layers with nonlinear activation function as in [46]. Assuming a deep neural network with $L$ layers, the output of the $l$-th layer of the nonlinear part of the first branch can be written as follows:

$$\Phi_e^{(l)} = \sigma\left(W_e^{(l)^T}\Phi_e^{(l-1)}\right), \quad l = 1, \dots, L, \quad (7)$$

with $\Phi_e^{(0)} = E^T$, where $(.)^T$ denotes the transposition of the matrix. In Eq. (6), $W_e^{(l)}$ denotes the weights used in the $l$-th layer, and $\sigma(.)$ is a nonlinear activation function such as sigmoid, hyperbolic tangent, and rectified linear unit (ReLU) functions. Nodes in each layer are numbered in an order of $K < h_1 < h_2 \dots < N$.

It is worth mentioning that bias terms continuously influence pixel composition and act as one of the endmembers observed from the data in the optimization process. As a result, bias terms violate the important property of endmembers, namely, no material can always be expected to appear in all pixel compositions unless a scene is made up of just one type of material [47]. This is why, although bias terms are important parameters in neural network architectures because they threshold the responses through activation functions and determine the impact of parameters for next layers, they are removed in q. (6). Apart from this, by doing so, the simplex set assumption - that is - an affine projection, upon which conventional methods are based, is kept [47].

Therefore, the nonlinear term in Eq. (5) can be approximate with $\boldsymbol{\Phi}_e^{(L)}$. Now, in order to estimate the linear part of the Eq. (5), the weights of the nonlinear part are restricted to make a linear combination of the endmembers. To do so, the transposition of the multiplication of the weights of the network is considered as the abundance fraction matrix $\boldsymbol{A}$. Since the number of endmembers present each mixed pixel is much less than the total number of endmembers, a sparseness constraint on the abundance fractions matrix can be applied [9]. In order to apply this constraint, instead of explicit sparsity regularization, we a soft thresholding ReLU activation function for the multiplications of the weights, thereby applying both sparseness constraint and ANC. Thus, the output of the linear part can be formulated as follows:

$$\boldsymbol{\mathcal{L}}_e = ReLU\left(\prod_{l=1}^{L}(\boldsymbol{W}_e^{(l)})\right)^T \boldsymbol{E}^T. \qquad (8)$$

Accordingly, the output of the first branch, a reconstruction of the rows of the hyperspectral data, is defined as follows:

$$\widehat{\boldsymbol{X}}_e = \boldsymbol{\mathcal{L}}_e + \boldsymbol{\Phi}_e^{(L)}. \qquad (9)$$

The following objective function, which involves the reconstruction loss between the output of the network and the hyperspectral data, is used to train the network to estimate the endmember matrix $\boldsymbol{E}$ and network parameters.

$$\mathcal{J}_e = \min_{E,W_e} \frac{1}{2P}\sum_{i=1}^{P}\|\boldsymbol{x}_i^r - \widehat{\boldsymbol{x}}_{e\_i}\|_F^2 \qquad (10)$$

where $\|.\|_F$ denotes the Frobenius norm and $\widehat{\boldsymbol{x}}_{e\_i}$ is the reconstruction of the $i$-th row of the data ($i$-th column of $\widehat{\boldsymbol{X}}_e$).

So as to apply ASC on the abundance fraction matrix, the method in [48] is adopted. In this method, the observation matrix $\boldsymbol{X}$ and endmember matrix $\boldsymbol{E}$ are augmented by a row of constants. In our method, we replace $\boldsymbol{X}, \boldsymbol{E}$, and $\boldsymbol{\Phi}_e^{(L)}$ with the following augmented matrices in the optimization process.

$$\widetilde{\boldsymbol{X}} = \begin{pmatrix} \boldsymbol{X} \\ \delta \boldsymbol{1} \end{pmatrix} \qquad (11)$$

$$\widetilde{\boldsymbol{E}} = \begin{pmatrix} \boldsymbol{E} \\ \delta \boldsymbol{1} \end{pmatrix} \qquad (12)$$

$$\widetilde{\boldsymbol{\Phi}}_e^{(L)} = \begin{pmatrix} \boldsymbol{\Phi}_e^{(L)} & \boldsymbol{0}^T \end{pmatrix}, \qquad (13)$$

where $\delta$ is a parameter balancing the impact of ASC. $\boldsymbol{1}$ and $\boldsymbol{0}$ are row vectors with all elements equal to one and zero, respectively. By employing the above equations in Eq. (9), the sum of the abundances at each pixel, i.e. the sum of each column's entries of $ReLU\left(\prod_{l=1}^{L}(\boldsymbol{W}_e^{(l)})\right)$, is forced to be one.

Fig. 1 illustrated the structure of proposed neural network. By training the model, proper weights and latent representations, i.e. endmembers, are learned from the data to make a well-balanced combination between linear and linear relations.

*B. The estimation of abundance fractions*

Similarly, the output of the second branch is the reconstruction of the columns of the data. These reconstructions are a combination of linear and nonlinear transpositions of the unknown latent variables $\boldsymbol{A}$. The output of the $l$-th layer of the nonlinear part of the second branch is formulated as follows:

$$\boldsymbol{\Phi}_a^{(l)} = \sigma\left(\boldsymbol{W}_a^{(l)T}\boldsymbol{\Phi}_a^{(l-1)}\right), \quad l = 1, \dots, L, \qquad (14)$$

Where $\boldsymbol{\Phi}_a^{(0)} = \boldsymbol{A}$ and $\boldsymbol{W}_a^{(l)}$ denotes the weights used in the $l$-th layer. Nodes in each layer are numbered in an order of $K < h_1 < h_2 \dots < P$. The nonlinear term in Eq. (6) is approximated with the output of the nonlinear part, $\boldsymbol{\Phi}_a^{(L)}$, and the linear part in Eq. (6) is approximated as follows:

$$\boldsymbol{\mathcal{L}}_a = \left|\left(\prod_{l=1}^{L}(\boldsymbol{W}_a^{(l)})\right)^T\right| ReLU(\boldsymbol{A}) \qquad (15)$$

Therefore, the reconstruction of the columns of the data is given by

$$\widehat{\boldsymbol{X}}_a = \boldsymbol{\mathcal{L}}_a + \boldsymbol{\Phi}_a^{(L)}. \qquad (16)$$

Through solving the following objective function, the abundance fraction matrix $\boldsymbol{A}$ and the second branch's network parameters are optimized.

$$\mathcal{J}_a = \min_{A,W_a} \frac{1}{2N}\sum_{i=1}^{N}\|\boldsymbol{x}_i^c - \widehat{\boldsymbol{x}}_{a\_i}\|_F^2 \qquad (17)$$

where $\widehat{\boldsymbol{x}}_{a\_i}$ is the reconstruction of the $i$-th column of the data ($i$-th column of $\widehat{\boldsymbol{X}}_a$).

Here, in order to impose ASC on the abundance fraction matrix, the same approach as in the first branch is applied. Therefore, the data matrix, $\boldsymbol{\Phi}_a^{(L)}$, and $\left|\left(\prod_{l=1}^{L}(\boldsymbol{W}_a^{(l)})\right)^T\right|$ are augmented as in Eq. (11), $\begin{pmatrix}\boldsymbol{E}\boldsymbol{\Phi}_a^{(L)}\\ \delta\boldsymbol{0}\end{pmatrix}$, and $\begin{pmatrix}\left|\left(\prod_{l=1}^{L}(\boldsymbol{W}_a^{(l)})\right)^T\right|\\ \delta\boldsymbol{1}\end{pmatrix}$, respectively.

Now, we have two branches working independently to estimate the endmembers and abundance fractions with a nonlinear model. In order to force these two independent branches to work together, by employing multi-task learning principles, we use the learned latent variables and introduce an auxiliary task that improves the performance of the two branches.

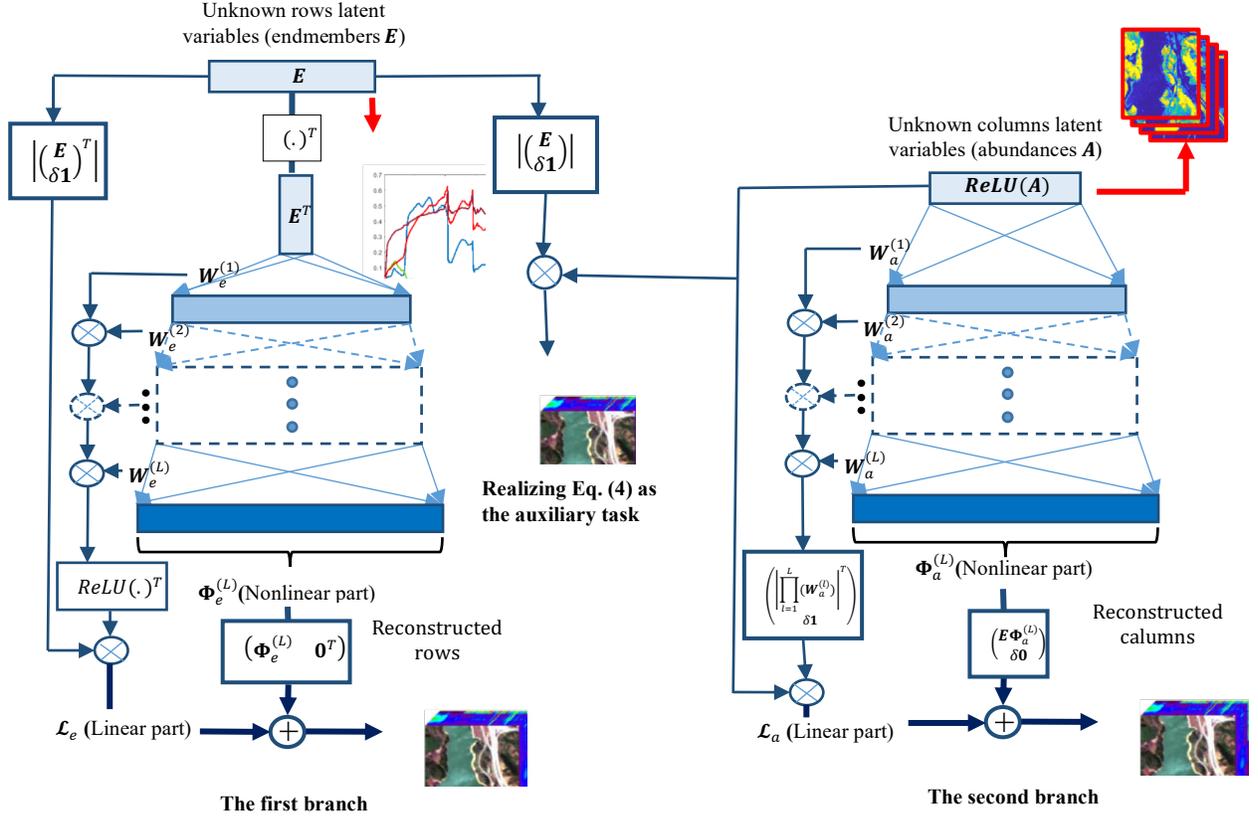

Fig. 2: The diagram of the proposed network. Endmembers are obtained from the first branch, abundances are obtained from the second branch, and the auxiliary task regularizes the endmembers and abundances.

## C. Regularizing $E$ and $A$

In multi-task learning, the information extracted by other related tasks are employed to improve the performance of the model's main task [49-51]. The process of multi-task learning often involves introducing one or more auxiliary tasks and sharing representations between all tasks. Using the information obtained from such auxiliary tasks makes an inductive transformation mechanism, improving the model through providing an inductive bias. This inductive bias drives the model to prefer hypotheses which account for more than one task, causing the solutions generalizing better for the main task. As a result, multi-task learning can be considered as a regularizer by introducing an inductive bias. Therefore, it mitigates the risk of overfitting and the Rademacher complexity of the model - the ability of the model to fit random noise [49].

With this motivation, the latent representations, $E$ and $A$, are forced to realize the model in Eq. (4). Therefore, the following loss function is introduced as an auxiliary task.

$$\mathcal{J}_m = \min_{E,A} \frac{1}{2N} \|X - EA\|_F^2 \qquad (18)$$

It is worth mention that Eq. (11) and (12) are applied to impose ASC. By employing the above equation in training process, the two branches are forced to work together. The complete structure of the proposed approach is illustrated in Fig. 2.

Apart from this, in order to drive the weights and inputs of the network to decay as a complementary to the multi-task learning regularizer, we use the Frobenius norm of weights and inputs, written as the following equation, as a regularization term.

$$\mathcal{J}_r = \frac{1}{2P}\|E\|_F^2 + \frac{1}{2N}\|A\|_F^2 + \frac{1}{2}\sum_{l=1}^{L}\|W_e^{(l)}\|_F^2 \\ + \frac{1}{2}\sum_{l=1}^{L}\|W_c^{(l)}\|_F^2 \qquad (19)$$

## D. Objective function and optimization

Multi-task learning involves a multi-loss framework. Several loss functions with different theoretical motivations improve the generalization capability of the network. In our model, the objective function consists of several components as follows:

$$\mathcal{J} = \mathcal{J}_e + \mathcal{J}_a + \alpha \mathcal{J}_m + \beta \mathcal{J}_r \qquad (20)$$

where $\alpha$ and $\beta$ are trade-off parameters controlling the importance of two regularization terms.

The objective function in Eq. (20) is non-convex; therefore, in order to obtain a suboptimal solution at local minima, such nonlinear optimization approaches as BFGS, LBFGS, iRprop+, etc. should be selected. Furthermore, since the inputs of the network, $E$ and $A$, are unknown, the selected optimization approach should be batch-wise rather than mini-batch or stochastic. Nevertheless, in large-scale datasets, a batch-wise LBFGS fails to scale properly with the number of examples, and as a result, to ensure a fast optimization method, a mini-batch LBFGS should be selected [52]. In such datasets, iRprop+ has shown its superiority over other optimization such as BFGS [53]. In addition, the computational cost of BFGS is more than that of iRprop+ [46], which can be a major drawback in large-scale datasets. Since hyperspectral data are large-scale, iRprop+ is selected as the optimization approach in our model.

The gradient of the objective function needs to be computed in the iRprop+ optimization approach. In order to compute the gradient of the objective function, $\partial \mathcal{J}_e/\partial E$, $\partial \mathcal{J}_e/\partial W_e$, $\partial \mathcal{J}_a/\partial A$, $\partial \mathcal{J}_a/\partial W_a$, $\partial \mathcal{J}_m/\partial E$, and $\partial \mathcal{J}_m/\partial A$ are obtained by back-propagation algorithm. We initialize the network weights as in [54], and the results of VCA [55] are used to initialize the endmember matrix $E$. It should be mentioned that, although random initialization can also be applied, VCA initialization, which provides more robust results, is common in hyperspectral unmixing algorithms [9, 13, 18]. To initialize abundance matrix $E$ the following equation is used.

$$A = E^{-1} * X. \qquad (21)$$

where $E^{-1}$ is the inverse of $E$, and for $A$ to be nonnegative the following equation is applied.

$$A = ReLU(A). \qquad (22)$$

It is worth mentioning that the range of the output of the network and the data should be equivalent. For evaluation, the data can first be transformed into an appropriate interval, and then after optimization, it can be returned to its original interval.

## IV. EXPERIMENT RESULTS

In this section, in order to evaluate the performance of the proposed method, several experiments have been conducted on synthesized and real-world datasets. The proposed method is compared with several conventional and deep learning-based unmixing approaches. The first approach is the robust nonnegative matrix factorization (rNMF) [18] which is a nonlinear unmixing algorithm estimating the endmembers and abundances by using a block-coordinate descent algorithm. In the second one, the endmembers are extracted employing N-finder (N-FINDR) [56] and the abundances are estimated applying a nonlinear unmixing method which uses variable splitting and augmented Lagrangian (NUSAL) [57]. The linear unmixing approach in [58], stacked nonnegative sparse autoencoders (SNSA), is the third approach. In SNSA, a set of stacked nonnegative sparse autoencoders is applied to find the outliers of the data, and then, an autoencoder is employed to estimate the endmembers and the abundance fractions. Addressing the post-nonlinear mixing problem, the autoencoder network in [59], which performs blind nonlinear unmixing, is the fourth approach. It will be referred to as the nonlinear autoencoder-based unmixing (NLAEU). The fifth approach is the deep autoencoder-based unmixing (DAEN) [38] performing linear blind unmixing by employing stacked and variational autoencoders. Abundance constrained multi-layer kernel non-negative matrix factorization (AC-MLKNMF) [60], a nonlinear unmixing method, is the last state-of-the-art approach whose results are compared with those of our method.

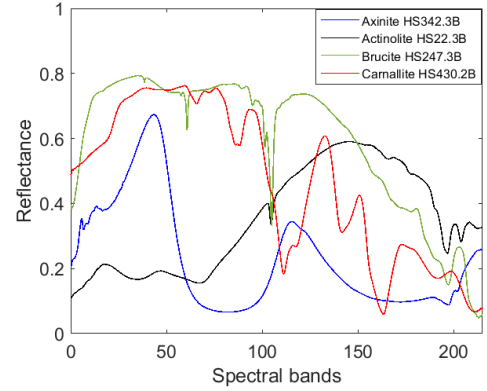

Fig. 3: Selected endmembers from USGS library

The same VCA results are used to initialize rNMF, SNSA, AC-MLKNMF, and our method. It should be noted that these methods are all adjusted according to the suggestions of the authors or are carefully set to achieve the best performance. ReLU is used for the activation function in the layers of our method, trade-off parameters $\alpha$ and $\beta$ are selected from {0.001, 0.005, 0.01, 0.05, 0.1, 0.5}, and $\delta$ is set to 5.

In order to measure the similarity between the real and estimated endmembers, the spectral angle distance (SAD) metric by the following definition is applied.

$$SAD = \frac{1}{K}\sum_{k=1}^{K} arc\ cos\left(\frac{e_k^T.\hat{e}_k}{\|e_k\|_2.\|\hat{e}_k\|_2}\right), \qquad (23)$$

where $e_k$ and $\hat{e}_k$ are the $k$-th real and estimated endmember, respectively. The accuracy of the abundances estimations is evaluated by the overall root mean square error (aRMSE) as follows:

$$aRMSE = \sqrt{\frac{1}{NK}\sum_{i=1}^{N}\|a_i - \hat{a}_i\|^2}, \qquad (24)$$

where $a_i$ and $\hat{a}_i$ indicate the true and estimated abundances of the $i$-th pixel, respectively.

### A. Experiments with synthetic data

In order to generate the synthetic data, we have randomly selected four signatures form Spectral signatures from USGS library (splib06) [61] as shown in Fig. 3. These spectral signatures contain 224 contiguous bands covering 380 nm to 2500 nm wavelength range. The data is generated with the LMM, the bilinear mixture model, and the post-nonlinear mixing model (PNMM) as described in [62]. For the LMM, Eq. (4) is used; the bilinear mixture model and PNMM are respectively given by the following equations:

TABLE I
RMSE RESULTS OF DIFFERENT METHODS FOR THE SYNTHETIC DATA

| | SNR=10dB | | | SNR=20dB | | | SNR=30dB | | |
|---|---|---|---|---|---|---|---|---|---|
| | LMM | bilinear | PNMM | LMM | bilinear | PNMM | LMM | bilinear | PNMM |
| rNMF | 0.1685 | 0.1524 | 0.1512 | 0.0656 | 0.0649 | 0.0676 | 0.0549 | 0.0598 | 0.0603 |
| N-FINDR-NUSAL | 0.1092 | 0.1223 | 0.1021 | 0.0398 | 0.0769 | 0.1021 | 0.0296 | 0.0495 | 0.1031 |
| SNSA | 0.0861 | 0.1597 | 0.1589 | 0.0471 | 0.0801 | 0.1037 | 0.0412 | 0.0676 | 0.1109 |
| NLAEU | 0.0812 | 0.0797 | 0.0694 | 0.0492 | 0.0627 | 0.0406 | 0.0384 | 0.0671 | 0.0349 |
| DAEN | 0.0843 | 0.1549 | 0.1596 | 0.0419 | 0.0781 | 0.1026 | 0.0388 | 0.0698 | 0.1131 |
| Proposed | **0.0586** | **0.0731** | **0.0548** | **0.0376** | **0.0573** | **0.0371** | **0.0263** | **0.0421** | **0.0297** |

Boldface numbers show the lowest RMSEs

TABLE II
SAD RESULTS OF DIFFERENT METHODS FOR THE SYNTHETIC DATA

| | SNR=10dB | | | SNR=20dB | | | SNR=30dB | | |
|---|---|---|---|---|---|---|---|---|---|
| | LMM | bilinear | PNMM | LMM | bilinear | PNMM | LMM | bilinear | PNMM |
| rNMF | 4.236 | 4.5768 | 6.9745 | 3.0921 | 4.2648 | 5.9863 | 3.2145 | 2.7895 | 5.4798 |
| N-FINDR-NUSAL | 19.2836 | 19.9215 | 18.9024 | 6.6232 | 6.7781 | 7.6682 | 1.9659 | 2.6156 | 5.9833 |
| SNSA | 4.0681 | 4.5963 | 6.6763 | 1.7631 | 3.9635 | 5.7146 | 0.6049 | 3.5968 | 5.4112 |
| NLAEU | 3.9235 | 4.5893 | 5.3146 | 1.1129 | 3.2369 | **5.4498** | 0.5862 | 3.4137 | 5.2453 |
| DAEN | 4.0012 | 4.7796 | 6.5469 | 1.0936 | 4.0593 | 5.8654 | **0.5324** | 3.8145 | 5.3661 |
| Proposed | **3.7723** | **4.2989** | **5.2136** | **0.9956** | **3.0912** | 5.4867 | 0.5584 | **2.5496** | **5.1821** |

Boldface numbers show the lowest SADs

TABLE III
SAD AND RMSE OF THE PROPOSED METHOD FOR SYNTHETIC DATASET, WITH DIFFERENT NUMBER OF ENDMEMBERS

| | Number of endmember | | | | | | | |
|---|---|---|---|---|---|---|---|---|
| | 3 | | 4 | | 6 | | 8 | |
| | bilinear | PNMM | bilinear | PNMM | bilinear | PNMM | bilinear | PNMM |
| SAD | 2.7678 | 4.5564 | 3.0912 | 5.4867 | 3.6842 | 5.9105 | 4.4622 | 6.6209 |
| RMSE | 0.0438 | 0.02436 | 0.0573 | 0.0371 | 0.0593 | 0.0404 | 0.0669 | 0.0476 |

$$x = Ea + \sum_{i=1}^{K-1}\sum_{j=1}^{K} a_i a_j (e_i \odot e_j) + n \qquad (25)$$

$$x = Ea + Ea \odot Ea + n, \qquad (26)$$

where $\odot$ indicates the Hadamard matrix product. The abundance fractions are generated from a Dirichlet distribution on the simplex which is defined by the ANC and ASC. Sensor noise and possible measurement errors are simulated with a zero-mean Gaussian noise whose signal-to-noise ratio (SNR) is set to 10, 20, and 30 dB, respectively. Three image with $256 \times 256$ pixels are generated for the experiments. In order to learn nonlinear interactions among the spectrum of the image, we use two hidden layers, because high-order interactions, if any are presented, are usually not very noticeable [62]. The number of nodes in the input layer, hidden layers, and output layer are set to [4 100 250 $N$] and [4 25 50 $P$] in the first and second branches, respectively.

The performance comparison in terms of RMSE and SAD of the LMM, the bilinear mixture model, and PNMM under different SNR for each method is shown in Table I and Table II, respectively. It can be seen that the proposed method provides the best abundance estimation results and outperforms other algorithms in endmember estimation performance in both linear and nonlinear models. Although SNSA and DAEN are based on the linear mixing model and their performance is not as good as other algorithms in abundance estimation in nonlinear cases, they are superior at detecting endmembers. It seems that deep learning-based algorithms demonstrate more effective blind unmixing performance in general. By comparing our method with state-of-the-art nonlinear mixing models – rNMF, NUSAL, and NLAEU – the accuracy of abundance estimation is nearly always improved in our method. Since

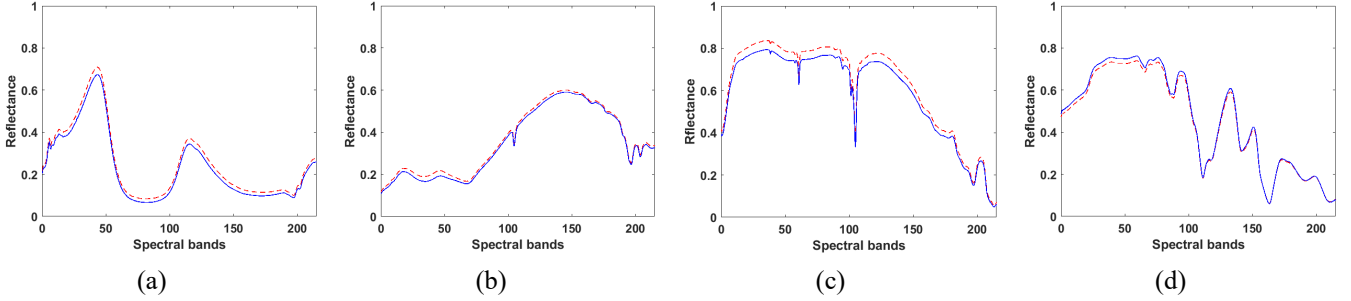

Fig. 4: Ground truth endmembers (blue solid lines) and estimated endmembers using the proposed method (red dashed lines) with the linear model under SNR=10dB. (a) Axinite HS342.3B, (b) Actinolite HS22.3B, (c) Brucite HS247.3B, and (d) Carnallite HS430.2B.

TABLE IV
COMPARISON OF SAD AND RMSE OF DIFFERENT METHODS ON JASPER RIDGE DATASET

| Method | RMSE | Tree | Water | Soil | Road | Mean SAD |
|---|---|---|---|---|---|---|
| rNMF | 0.1398 | 0.1826 | 0.1388 | 0.2541 | 0.1064 | 0.1704 |
| N-FINDR-NUSAL | 0.1364 | 0.1559 | 0.1337 | 0.2253 | 0.0906 | 0.1514 |
| SNSA | 0.1059 | 0.1679 | 0.1263 | 0.2181 | 0.0909 | 0.1508 |
| NLAEU | 0.0904 | 0.0754 | 0.1186 | 0.1021 | 0.0597 | 0.0889 |
| DAEN | 0.0957 | 0.1774 | 0.3237 | 0.1123 | **0.0588** | 0.1680 |
| AC-MLKNMF | 0.1300 | **0.0628** | 0.1292 | 0.0938 | 0.0641 | 0.0875 |
| Proposed | **0.0873** | 0.0695 | **0.1091** | 0.0921 | 0.0594 | **0.0825** |

Boldface numbers show the lowest SADs and RMSEs

TABLE V
COMPARISON OF SAD AND RMSE OF THE PROPOSED METHOD ON JASPER RIDGE DATASET WITH DIFFERENT VALUES OF $\alpha$

| | $\alpha$ | | | | | | |
|---|---|---|---|---|---|---|---|
| | 0 | 0.001 | 0.005 | 0.01 | 0.05 | 0.1 | 0.5 |
| SAD | 0.0911 | 0.0855 | 0.0849 | 0.084 | 0.0837 | 0.0829 | 0.0825 |
| RMSE | 0.1321 | 0.1193 | 0.1101 | 0.1074 | 0.1009 | 0.0941 | 0.0873 |

NUSAL is a kernel-based algorithms, the selection of the kernel and its parameters causes great effects on its performance. On the other hand, our model learns the nonlinearity from the data; thus avoiding the problem of kernel selection. In Fig. 4, the comparison between the endmembers extracted by the proposed method and the reference ones are shown. It can be seen from Fig. 4 that our method is capable of extracting endmembers hardly different with the references in the USGS spectral library.

Since the nonlinear interactions among the spectrum of the synthetic data become more and more complex with the increase in the number of endmembers, we evaluate the performance of the proposed method under different number of endmembers. The SAD and RMSE results of the two nonlinear models are listed in Table III. One can observe that the proposed method still provides sufficiently good results when the number of endmembers increase.

*B. Experiments with real datasets*

For further evaluation, two real datasets – Jasper Ridge dataset and Cuprite dataset – are utilized to verify the effectiveness of the proposed method.

*1) Jasper Ridge:*

This dataset is a hyperspectral image of a rural area at Jasper Ridge, California, USA, which is recorded by airborne visible / infrared imaging spectrometer (AVIRIS) sensor. It contains $512 \times 614$ pixels recorded at 224 wavelength bands covering $380\ nm$ to $2500\ nm$ wavelength range at a ground sampling distance (GSD) of $20\ m$. A widely-used subimage with size $100 \times 100$ pixels, in which noisy and water absorption bands are removed and 198 band are retained, is used in this set of experiments. In this subimage four endmembers are investigated: #1 Tree, #2 Water, #3 Soil, and #4 Road. The parameters of our network are set as in the previous part.

In Fig. 5, the endmembers obtained by implementing the proposed method on Jasper Ridge are visually compared with the ground truth endmembers. It can be seen that the proposed method is capable of providing accurate estimation of the endmembers in a real-world dataset. SAD values for each endmember and overall SAD and RMSE values for this dataset are listed in Table IV. The results in Table IV shows that AC-MLKNMF, because of adding $L_{1/2}$ sparsity constraint and piecewise smoothing constraint to the abundances and performing layered processing to fully decompose the data, provides relatively good performance. One can also observe that, in comparison with synthetic data, the deep learning-based unmixing approaches provide more accurate results in both RMSE and SAD values. In particular, it is clear that our method outperforms its competitor in both the abundance and endmembers estimation. In Fig. 6, a comparison of the compared method on the estimation of abundance maps is demonstrated. While, in some ways, DAEU and DAEN are incorrect in their estimation of the *Road* as *Water*, it is clear that deep learning-based approaches estimate the abundance in a realistic manner. We can see that the estimated abundance map of our method, in comparison with the other methods, has a lower deviation from the ground truth abundance map.

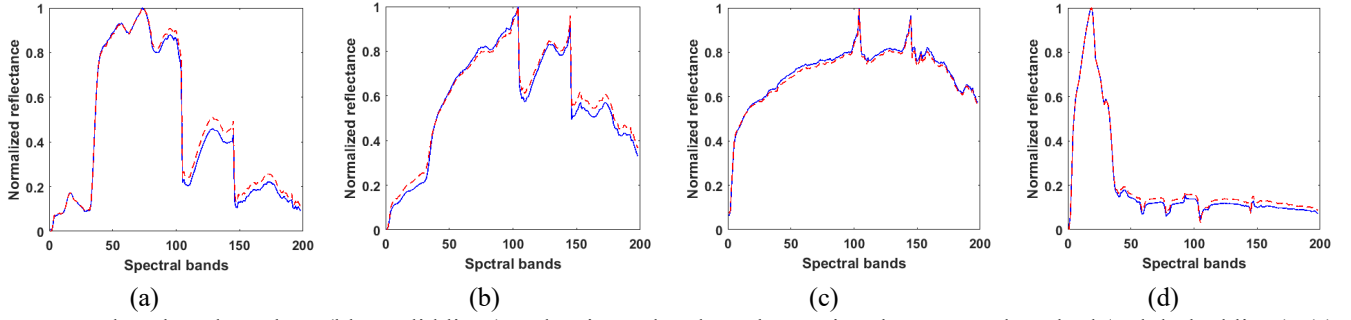

Fig. 5: Ground truth endmembers (blue solid lines) and estimated endmembers using the proposed method (red dashed lines). (a) Tree (b) Water, (c) Road, and (d) Soil.

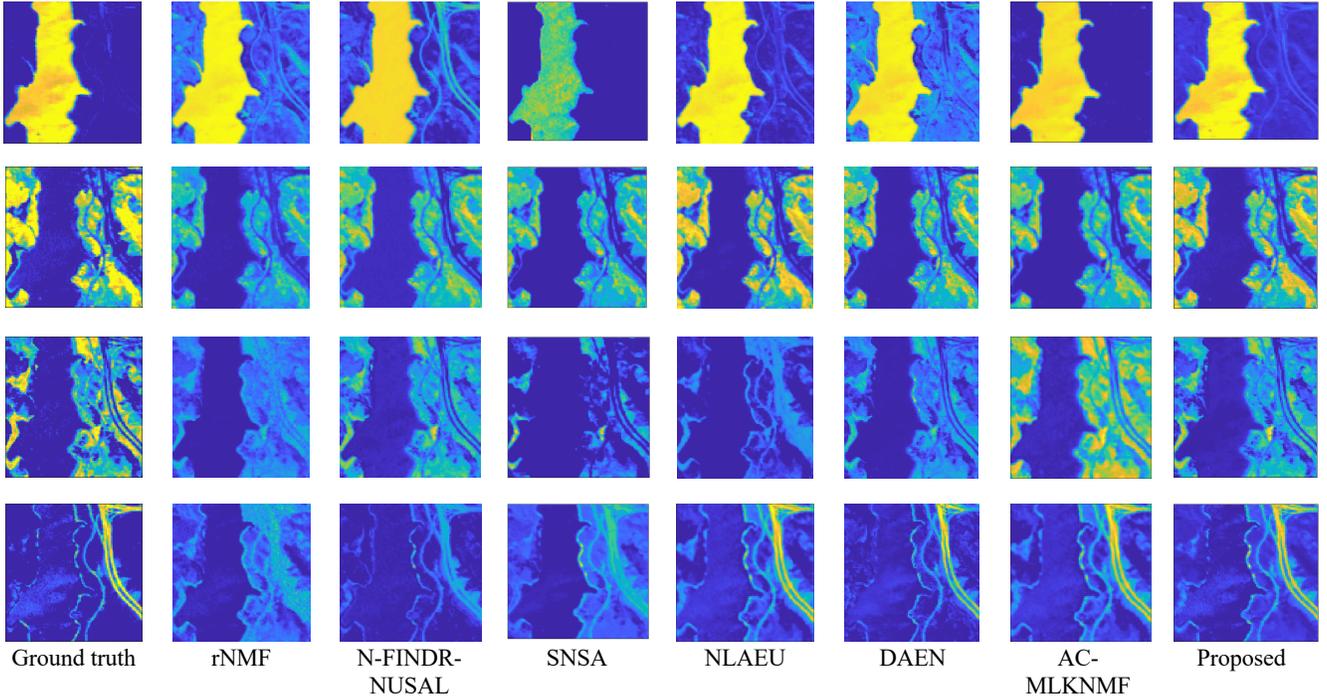

Fig. 6: Estimated abundance map on Jasper Ridge dataset obtained by different methods. From top to bottom: Water, Tree, Soil, and Road.

In order to evaluate the influence of the multi-task learning on the estimation accuracy of the endmembers and the abundance fractions, we implement the proposed method on the Jasper Ridge dataset with different values of $\alpha$. Overall RMSE and SAD values for different $\alpha$ are reported in Table V. We can see that, when the two branches works separately, i.e. $\alpha = 0$, the network does not perform well in both the abundances and endmember estimation. By forcing the two branches to work together and applying multi-task learning, i.e. $\alpha \neq 0$, the performance of the both branches improves noticeably. The large values of $\alpha$ shows the importance of the objective function of the auxiliary task, which is based on the linear mixing model.

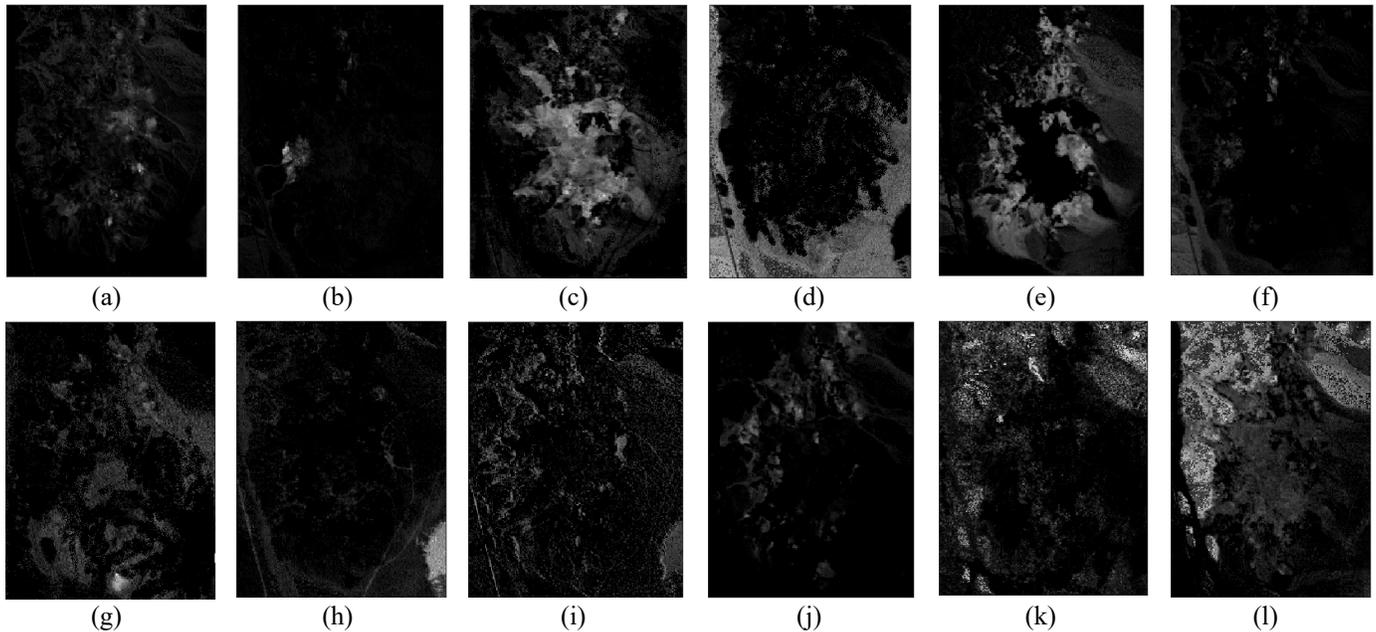

Fig. 7: Estimated abundance map on Cuprite dataset obtained by the proposed method. (a) Alunite, (b) Buddingtonite, (c) Chalcedony, (d) Jarosite, (e) Kaolin/Smect, (f) Kaolinite wx1, (g) Kaolinite px1, (h) Montmorillonite, (i) Muscovite, (j) Nontronite, (k) Pyrope, and (l) Sphene.

TABLE VI
COMPARISON OF SAD RESULTS OF DIFFERENT METHODS ON CUPRITE DATASET

| Method | rNMF | N-FINDR-NUSAL | SNSA | NLAEU | DAEN | AC-MLKNMF | Proposed |
|---|---|---|---|---|---|---|---|
| Alunite | **0.0846** | 0.0992 | 0.1201 | 0.1237 | 0.1097 | 0.1282 | 0.1121 |
| Buddingtonite | 0.1364 | 0.1639 | 0.1062 | 0.1103 | **0.0939** | 0.1092 | 0.1102 |
| Chalcedony | 0.1439 | 0.1687 | 0.1097 | 0.0974 | 0.1104 | **0.0568** | 0.0945 |
| Andradite | 0.1447 | 0.1893 | 0.1131 | 0.0701 | 0.1095 | 0.0743 | **0.0701** |
| Kaolinite_1 | 0.0928 | 0.0804 | 0.0819 | 0.0697 | 0.0801 | 0.0567 | **0.0549** |
| Kaolinite_2 | 0.0841 | 0.0792 | 0.0794 | **0.0748** | 0.0786 | 0.0797 | 0.0814 |
| Dumortierite | 0.0214 | **0.0108** | 0.1079 | 0.1124 | 0.1086 | 0.2029 | 0.1689 |
| Montmorillonite | 0.0604 | 0.0651 | 0.0616 | 0.0746 | **0.0583** | 0.0771 | 0.0756 |
| Muscovite | 0.1221 | 0.1437 | 0.1252 | 0.1244 | 0.1303 | **0.1234** | 0.1288 |
| Nontronite | 0.0704 | 0.0803 | 0.0825 | 0.0803 | 0.0796 | 0.0744 | **0.0719** |
| Pyrope | 0.1097 | **0.0532** | 0.0638 | 0.0916 | 0.0593 | 0.1055 | 0.1077 |
| Sphene | 0.1416 | 0.1496 | 0.1396 | 0.1264 | 0.1407 | 0.0687 | **0.0677** |
| Mean SAD | 0.1010 | 0.1070 | 0.0993 | 0.0963 | 0.0966 | 0.0964 | **0.0953** |

Boldface numbers show the lowest SADs.

The best results are obtained when $\alpha = 0.5$, which indicates that the auxiliary task plays an important role in the unmixing process. One possible reason is that, since the transitional or boundary zones between different regions – such as at the shoreline of the water – are more likely to have nonlinear interactions, the number of pixels corresponding to such regions is less than the number of pixels corresponding to regions for which the linear mixing model is applicable. Therefore, the linear mixing model has an important role in the unmixing process. It is worth mentioning that, by choosing ReLU as the activation function of the last layer of the network, the sparsity of the nonlinear term can be applied.

*2) Cuprite:*
This dataset is the hyperspectral image of the Cuprite mineral area in Nevada, USA, collected by AVIRIS sensor. It contains $250 \times 191$ pixels recorded at 224 wavelength bands. After removing noisy and water absorption bands (1-2, 104-113, 148-167, 221-224), 188 bands remains as the second real dataset. In this dataset, 12 endmembers are investigated [9], and because ground truth information of the abundances are not available for a performance evaluation, only SAD values of the state-of-the-art algorithms are compared. These results are reported in Table VI. In addition, the grayscale abundance maps estimated by the proposed method are illustrated in Fig. 7. Because of the large number of endmember existing in this dataset, Cuprite is a complex dataset which leads to a more challenging unmixing process than that of the Jasper Ridge dataset. However, as the results in Table VI and the abundance maps in Fig. 7 shows, our method is still capable of providing sufficiently good results for both the abundances and endmembers estimation.

## V. Conclusion

This paper presents a deep learning-based approach for unsupervised nonlinear unmixing. The proposed method consists of a linear mixture component and an additive nonlinear mixture component, which enables the model to learn the inherent nonlinearity from the data by applying the modeling potential of deep neural networks. In addition, our method benefits from multi-task learning principles, which improves the accuracy of both abundance fractions and endmembers estimation. By conducting several experiments on synthetic and real datasets and comparing the results of the proposed method with the results of six state-of-the-art unmixing methods, in both the abundances and endmembers estimation, the superiority of our method is shown. As the future work, further improvements will be made to the proposed network by integrating the spectral-spatial information of the hyperspectral image into the proposed network.